\documentclass[aps,prb,twocolumn,groupedaddress,showpacs]{revtex4}

\usepackage{graphicx}

\begin{document}


\title{Diameter dependence of ferromagnetic spin moment in Au nanocrystals.}


\author{H. Hori, Y. Yamamoto, T. Iwamoto, T. Miura, T. Teranishi and M. Miyake}
\affiliation{School of Materials Science, Japan Advanced Institute of Science and Technology (JAIST), 1-1 Asahidai, Tatsunokuchi, Ishikawa, 923-1292, Japan}


\date{\today}

\begin{abstract}
Au nanoparticles exhibit ferromagnetic spin polarization and show diameter dependence in magnetization. The magnetic moment per Au atom in the particle attains its maximum value at a diameter of about 3 nanometer (nm) in the Magnetization-Diameter curve. Because Au metal is a typical diamagnetic material, its ferromagnetic polarization mechanism is thought to be quite different from the ferromagnetism observed in transition metals. The size effect strongly suggests the existence of some spin correlation effect at the nanoscale. The so-called ``Fermi hole effect'' is the most probable one given in the free electron gas system. Ferromagnetism in Au nanoparticles is discussed using this model.
\end{abstract}


\maketitle

\section{Introduction}
Bulk Au metal is chemically stable and has diamagnetic properties. However, Au particles show ferromagnetism at the nanoscale(nm) \cite{Teranishi1,HoriPLA,HoriNATO,Li}. Such ferromagnetic spin polarization is surprisingly unique and the polarization mechanism is quite interesting. It should be emphasized that these magnetic effects are important phenomena that emerge only at the nanoscale. The Au atom has the electron configuration [Xe-core] $(4f)^{14}(5d)^{10}(6s)^{1}$. Application of the electron gas model to the 6$s$ electrons in bulk Au metal is considered appropriate because the $d$-band is deep enough below the Fermi level and $s$-$d$ hybridization is negligibly small\cite{Ashcroft}. The molecular orbits in which electrons move around whole atoms in a nanoparticle correspond to the conduction band in the bulk state. The energy levels of the molecular orbits in a nanoparticle are discrete in a spatially narrow potential. These molecular orbits are believed to correspond to the conduction electron states in the bulk metal when energy gaps of less than $\sim 10$ K ($\sim 0.001$ eV) can be ignored. In the present work, the electron in a molecular orbit is referred to as a ``conduction electron'' in the nanoparticle.

For the last few decades, the problem of ferromagnetic polarization in a free electron gas has been discussed as a basic physical problem related to the Fermi hole effect. However, no ferromagnetism has been observed to date. In connection with the Fermi hole effect, ferromagnetic spin correlation near the surface has been discussed theoretically \cite{Harbola,Juretschike}. A unique characteristic of nanoparticles is the large proportion of surface atoms. In fact, the ratio is about 45 at\% at a diameter of 2.5 nm. Thus, we expect i) the emergence of spin polarization resulting from the Fermi hole effect ii) a diameter dependence of magnetization in the case nanoparticles. However, no theoretical studies have been conducted so far. It is against this background that the present study investigates the diameter dependence of magnetism in Au nanoparticles. The results are assessed under the assumption of surface ferromagnetism.

\section{Experimental procedure}

\subsection{Sample preparation and characterization}

Isolated nanoparticle are considered to be stable in free space. In a cluster, however, they spontaneously form large particles (diameter $> 10$ nm) to reduce their surface energy. Protective agents provide a practical means of preventing this cohesion. Much research has been devoted to developing protective agents for various nanoparticles. For Au nanoparticle, protective agents such as polyacrylonitrile (PAN), polyallyl amine hydrochloride (PAAHC), polyvinyl pyrolidone (PVP) and dodecane thiol (DT) have been reported \cite{Teranishi2}. The question of how these agents influence the electronic state of nanoparticles has largely been neglected except in the case of DT. Moreover, nanoparticles encapsulated in protective agents are magnetically isolated from each other, even in the case of DT. \cite{HoriPLA}.

\begin{figure}[htbp]
\includegraphics[width=7cm]{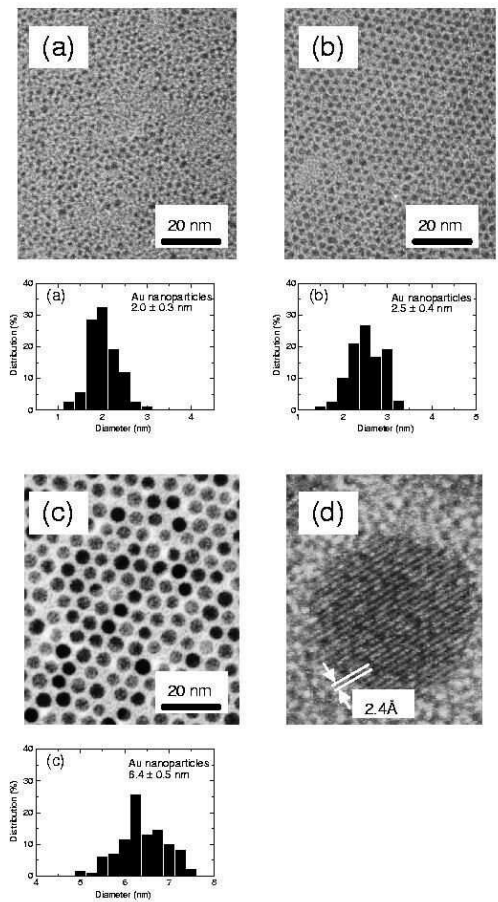}
\caption{\label{fig:TEM} TEM image of Au nanoparticles protected by DT and their diameter distributions for samples having an average diameter of (a) $2.0 \pm 0.3$, (b) $2.5 \pm 0.4$, (c) $3.5 \pm 0.5$ nm. (d) HRTEM image of sample (c). Scale bar is included in the image.}
\end{figure}
\begin{figure}[htbp]
\includegraphics[width=7cm]{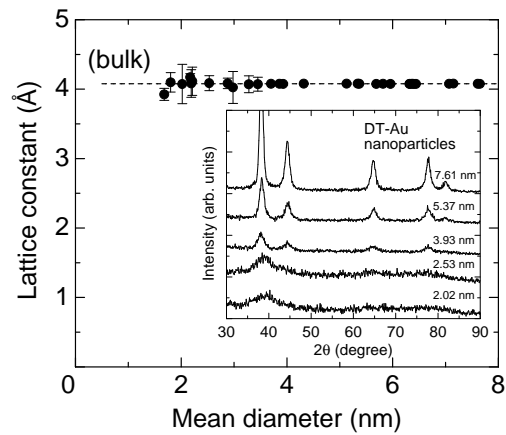}
\caption{\label{fig:XRD} Diameter dependence of lattice constants in nanoparticles. 
The lattice constants are estimated from the XRD data in the inset.}
\end{figure}

The diameter distribution of the sample was observed under a transmission electron microscope (TEM). Figure \ref{fig:TEM} shows a typical example of the diameter distributions observed in the present study. The crystalline layer structure of a Au nanoparticle is visible in the close-up image (d). The lattice plane spacing revealed in Fig. \ref{fig:TEM} corresponds to the fcc structure of the bulk metal. The estimated lattice spacing of 0.24 nm in the nanoparticles is consistent with that of the bulk metal and the X-ray diffraction data show that the lattice constant does not change, at least for diameters above 2.5 nm, as seen in Fig. \ref{fig:XRD}.

No edge distortion or deformation of the lattice layer is observed in the close-up image of the nanoparticle. This means that the nanoparticle exhibits a crystal structure consisting of a stack of lattice planes and that the crystal symmetry of the bulk Au metal is retained within the nanoparticle. The surface is highly important in nanoparticles: the surface potential has a major influence on the electron wave function. Indeed, what distinguishes nanoparticles from bulk metal are the surface-induced effects observed in the former. The anomalous ferromagnetic polarization observed in Au nanoparticles is presumably also related to these surface effects.

\begin{figure}[htbp]
\includegraphics[width=7cm]{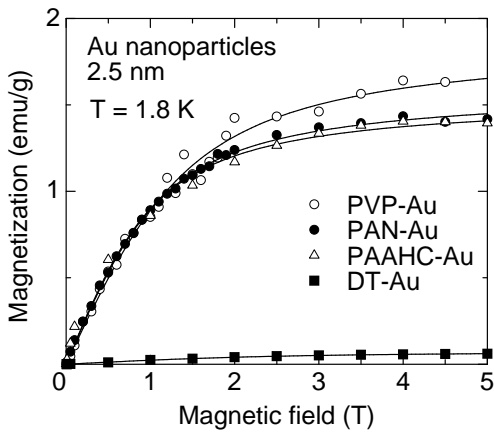}
\caption{\label{fig3} The effect of protective agents on magnetization of Au nanoparticles.}
\end{figure}

Figure \ref{fig3} shows the effect of protective agents on magnetization of Au nanoparticles with a mean diameter of 2.5 nm. It is noteworthy that the saturation magnetization is remarkably small in the case of DT. The most characteristic property of a Au nanoparticle in DT is the strong covalent bond between the thionic group and Au \cite{Teranishi2, Hostetler}. This covalent bond is presumed to cause the reduction in magnetization by inducing a spin-singlet state. The singlet state electrons are presumably localized around the interface between the Au surface and the sulfur in thionic groups, assuming that the distance dependence governing the formation of the singlet state is similar to the usual exchange interaction. Although nanoparticles in PAN, PAAHC and PVP form coordinate bonds and large reductions in magnetization are not observed, a slight difference of saturation moments in Fig. \ref{fig3} is noticeable. This difference may be explained by the presence of a weak covalent bond in addition to the main coordinate bond.

Despite the marked reduction in magnetization of Au nanoparticles in DT, the field and temperature dependencies are both consistent with those for other protective agents when comparing normalized saturation magnetizations. In the case of PAAHC, for example, the peak in the Magnetization-Diameter curve is observed at a diameter of 2.7 nm, although the data are scattered because of the broad diameter distribution. This peak corresponds to the peak observed at 3 nm in the case of DT. Because many Au atoms in the surface layer form a covalent bond in DT and generate localized spin singlet states, they modulate the surface potential. Thus, the new surface to the free electron system is formed just inside the covalent bonding region. On the other hand, the large reduction in saturation moment strongly supports the model of surface ferromagnetism in nanoparticles.

Despite the moment reduction, the protective agent DT has the following favorable properties: 1) Wide range diameter control from 1.5 to 10 nm. 2) Sharp diameter distribution is easily realized. 3) Despite the large modulation in surface potential, electron gas properties are retained within the interface region of the nanoparticle. 4) Almost all magnetic properties of Au nanoparticles in DT correspond to properties observed in other protective agents, provided the saturation magnetizations are normalized.

Thus, all samples in the present study were prepared using DT. These samples were used to investigate the diameter dependence of magnetization over a wide diameter range. Such an investigation is important in that it helps elucidate the origin of ferromagnetism in Au nanoparticles given that the diameter range in the present study is on the order of the Fermi hole diameter.

\subsection{Magnetization measurement}

Magnetization was measured by a SQUID magnetometer. The spontaneous spin polarization of Au atoms in Au nanoparticles has recently been confirmed using X-ray MCD \cite{Yamamoto}. These X-ray MCD measurements were performed in the SPring-8 facility in Japan. Low temperature experiments were conducted to estimate the saturation magnetization. If the Au nanoparticles exhibited diamagnetism or Pauli paramagnetism, the field dependence would have been a straight line. But in the original data, a nonlinear curve is superimposed on a negative linear dependence. The diamagnetic contribution of the protective agent (DT) and sample holder were previously measured. Subtracting  these contributions yields the magnetization curve. The curve can be represented by the Langevin function, which exhibits quasi-paramagnetic field and temperature dependences. This quasi-paramagnetism of nanoparticles is referred to as ``superparamagnetism''. Figure \ref{fig4} shows the diameter dependence of the magnetic moment of the nanoparticles. For samples with a largely asymmetric diameter distribution, the data points are fairly scattered. Thus, the present study used samples with a relatively symmetric distribution.

Original magnetization data obtained at liquid helium temperatures include a certain ``inherent linear field dependence of magnetization''. Because diamagnetism is temperature independent and superparamagnetism vanishes at high temperatures, a diamagnetic moment is observed at room temperature. Figure \ref{fig5} shows the diameter dependence of the inherent diamagnetic moment of the nanoparticle. 

\section{Results and Discussion}

\subsection{Results of magnetization measurement}

\begin{figure}[htbp]
\includegraphics[width=7cm]{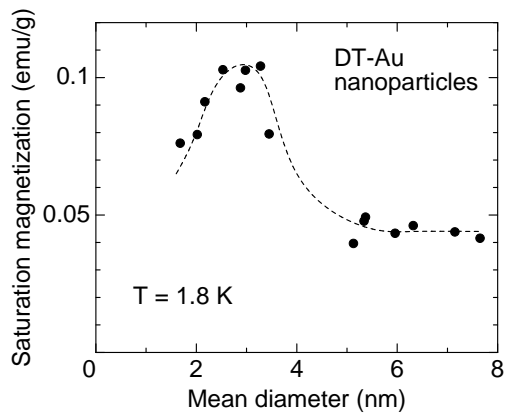}
\caption{\label{fig4} Magnetization as a function of Au nanoparticle diameter.  Magnetization is proportional to the magnetic moment per Au atom.}
\end{figure}

\begin{figure}[htbp]
\includegraphics[width=7cm]{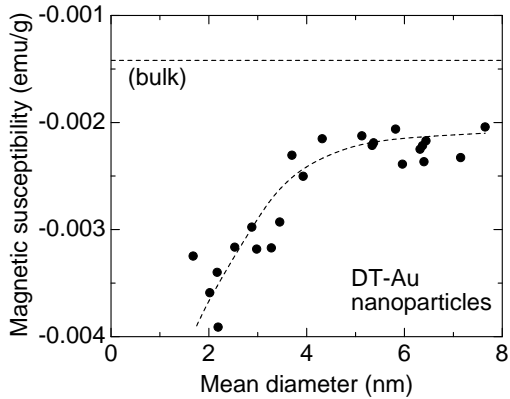}
\caption{\label{fig5} Diameter dependence of diamagnetism in Au nanoparticles.}
\end{figure}

The ferromagnetic moment of each nanoparticle can be estimated by the Langevin function, assuming a sharp diameter distribution. Although the magnetization data in Fig. \ref{fig4} are scattered, the peak at about 3 nm is clearly visible. The diamagnetic moment increases with decreasing diameter, as shown in Fig. \ref{fig5}. The typical equation for diamagnetic susceptibility $\chi_{\text{atom}}$ in the case of neutral atoms is:

\begin{equation}
\label{eq1}
\chi_{\text{atom}} = \frac{\mu_0 Z}{6m}\sum {< r^2 > } \quad ,
\end{equation}

\noindent
where $Z$, $<r^{2}>$, $m$ and $\mu_{0}$ are electron number, mean square of the electron orbit radius, mass of electron and magnetic permeability of vacuum, respectively. According to this equation, the diamagnetic susceptibility of the nanoparticle should increase with the electron orbit diameter, which is contradicted by the experimental results shown in Fig. \ref{fig5}.

\subsection{Possible model}

The Fermi hole effect and Hund's rule are two possible mechanisms whereby ferromagnetic spin polarization may arise in nanoparticles. The spin correlation of the ``Fermi hole effect'' near the surface region of a free electron gas system has been discussed \cite{Harbola}. Each electron spin in the Fermi hole state is spherically surrounded by oppositely directed spins and the diameter of the sphere is related to the Fermi wavelength of $1/k_{\text{F}}$ where $k_{\text{F}}$ is the Fermi wave number. Despite these spin correlations, up and down spins are totally cancelled in usual metals, resulting in zero net spin polarization. However, an imbalance may occur near the surface. Indeed, Okazaki and Teraoka have theoretically predicted the existence of ferromagnetism in thin films \cite{Okazaki}. An imbalance between the up and down spins is expected to emerge in the case of nanoparticles with a diameter smaller than the Fermi hole, however, there are no clear theoretical predictions so far. Such an imbalance would lead to ferromagnetic spin polarization. The spin correlation length from the surface is given by $\pi /k_{\text{F}}$ and the numerical value for Au is approximately 1.5 nm \cite{Harbola}.

Hund's rule is the origin of spin polarization in atoms and ions. The applicability of Hund's rule to the nanoparticle has been discussed for the diameter range of up to 5 nm by use of the Hatree-Fock approximation \cite{Asari}. The Fermi hole effect and Hund's rule appear to be independent concepts. However, in the case of nanoparticles, Hund's rule should be continuous with the Fermi hole effect for large diameters because both concepts are derived from the Pauli Principle as applied to the many electron system in the three-dimensional (3D) well-type potential. Thus, the only difference between the two concepts is that they apply at different size scales. Otherwise, they are consistent with one another: it is speculated that the outermost shell in the 3D well potential has spin polarization and the shell thickness is on the order of the Fermi wavelength. Moreover, the closed shell core might correspond to the inside of the nanoparticle.

\subsection{Diameter dependence of magnetization}

The magnetization data in Fig. \ref{fig4} is proportional to the magnetic moment per Au atom. The diameter at maximum magnetization (3 nm) is twice the surface spin correlation depth $\pi/k_{\text{F}}$. If Au atoms in the nanoparticle are uniformly magnetized, it is hard to find a clear and consistent explanation for the appearance of the Magnetization-Diameter curve. If the surface magnetism model is applied, the experimental data is easily explained under the following assumptions: 1) depth of spin-correlated surface region is also about $\pi/k_{\text{F}}$ in the nanoparticle and this region mainly contributes to the spin polarization. 2) At nanoparticle diameters larger than $\pi/k_{\text{F}}$, the spin-correlation region is limited to within the shell of surface atoms and the core of the particle contributes little to the ferromagnetism. These assumptions are justifiable based on the results presented above.

\begin{figure}[htbp]
\includegraphics[width=7cm]{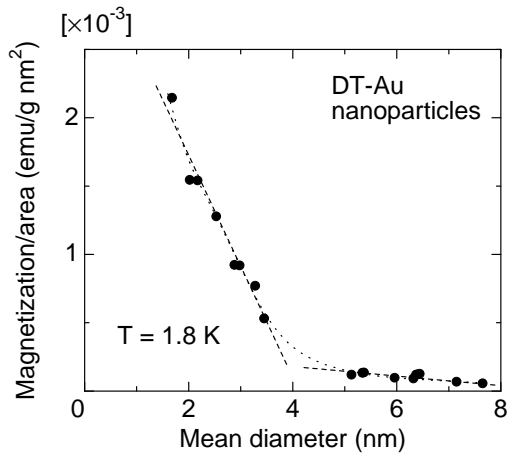}
\caption{\label{fig6} Magnetization per nanoparticle surface area. The data are calculated from Fig. \ref{fig4}}
\end{figure}

The data in Fig. \ref{fig6} show the diameter dependence of magnetization per nanoparticle surface area. Figure \ref{fig6} suggests that the degree of imbalance increases linearly with decreasing diameter. The figure also indicates that spin polarization virtually vanishes at a diameter of about 4 nm. Above this critical diameter, the moment of the Au nanoparticle is consistent with that of the bulk metal.

The diamagnetic component of Au nanoparticle shows anomalous diameter dependence as shown in Fig. \ref{fig5}. To validate the surface model for the ferromagnetism observed in Au nanoparticles, the model is used to analyze the diameter dependence of diamagnetism in Fig. \ref{fig5}.

Generally, the total susceptibility $\chi$ of bulk Au is theoretically given \cite{Okazaki} by:

\begin{equation}
\label{eq2}
\chi = \chi_{\text{Pauli}} + \chi_{\text{Band}} + \chi_{\text{ion}} \quad ;
\end{equation}

\begin{equation}
\chi_{\text{Pauli}} \cong \frac{3N\mu_{\text{B}}^2}{2\varepsilon_{\text{F}}} , \chi_{\text{Band}} \cong \frac{-N\mu_{\text{B}}^2}{2\varepsilon_{\text{F}}}\left({\frac{m}{m^\ast}} 
\right)^2
\end{equation}

\noindent
where $\chi_{\text{Band}}$, $\chi_{\text{ion}}$ and $\chi_{\text{Pauli}}$ are diamagnetic susceptibilities of the conduction electron and ion core and the Pauli paramagnetic susceptibility, respectively. $N$, $\mu_{\text{B}}$, $\varepsilon_{\text{F}}$, $m$ and $m^*$ are electron density, Bohr magneton, Fermi energy, electron mass and effective mass, respectively. These susceptibilities were numerically estimated using the standard theory \cite{Takahashi}. In particular, the diamagnetic susceptibility of \textit{$\chi $}$_{\text{ion}}$ of Au is larger than the Pauli paramagnetism in bulk state. Thus, \textit{$\chi $}$_{\text{ion}}$ mainly contributes to the diamagnetism of bulk Au metal and the diamagnetism is reduced by Pauli paramagnetism.

Pauli paramagnetism also reduces the diamagnetic moment of the core of nanoparticles. The surface diamagnetism, however, is not reduced, because spins corresponding to Pauli paramagnetism contribute to the ferromagnetism, manifesting themselves as the ferromagnetic component. Thus, the driving force for reduction in diamagnetism vanishes in small nanoparticles. According to Fig. \ref{fig6}, the net diamagnetic moment of a nanoparticle should increase with decreasing diameter. The result is qualitatively consistent with the model of the surface spin correlation based on the Fermi hole effect. The possibility of such a simple and clear interpretation of the experimental results strongly supports the model used in the present work.

\section{Conclusion}

The following anomalous experimental results were observed in Au nanoparticles: diameter dependence of ferromagnetism and an increase in diamagnetic moment with decreasing diameter. The Magnetization-Diameter curve exhibited a peak at a diameter of about 3 nm. The standard magnetization mechanisms fail to provide a consistent explanation for these nanoparticle-related phenomena. The Fermi hole effect borrowed from thin film theory, however, allows a consistent explanation. Thus, even though its applicability has yet to be confirmed, the mechanism of spin polarization based on the Fermi hole effect is probably at work in nanoparticles.

Once the surface ferromagnetism model is applied to nanoparticles, our seemingly anomalous experimental results can be fully explained. In particular, the peak in the Magnetization-Diameter curve can be accounted by the difference in diameter dependency of the decrease in surface magnetism and increase in spin polarized electron number. It should be noted that, at diameters larger than 4 nm, the magnetism reverts back to that of bulk state Au. This spin polarization mechanism is completely different from the ferromagnetism in transition metals with the spin correlation in the $d$-band. It is only at the nanoscale that the spatially restricted electron gas generates an imbalance in spin polarization. Thus, the ferromagnetism observed in the present work is thought to constitute a fundamental nanophysical phenomenon.

\begin{acknowledgments}
This work was supported in part by a grant from the Ministry of Education, 
Culture, Sports, Science and Technology, Government of Japan 
(MONBUKAGAKUSHO).
\end{acknowledgments}


\end{document}